\documentclass[conference]{IEEEtran}
\IEEEoverridecommandlockouts

\usepackage{cite}
\usepackage{bm}
\usepackage{amsmath}               
\usepackage{xpatch}
\usepackage{amsthm}
\usepackage{listings}
\usepackage{verbatim}
\usepackage{amssymb}
\usepackage{mathtools}
\usepackage{dsfont}
\usepackage{hyperref}
\usepackage{xcolor}
\usepackage{float}
\usepackage{lastpage}
\usepackage{fancyhdr}
\usepackage{relsize}
\usepackage{subfig}
\usepackage[nameinlink,capitalise]{cleveref}
\newtheorem{theorem}{Theorem}
\newtheorem{corollary}{Corollary}
\newtheorem{lemma}{Lemma}
\newtheorem{assumption}{Assumption}

\DeclareMathOperator*{\argmin}{arg\,min}
\hypersetup{
    colorlinks,
    linkcolor={red!0!black},
    citecolor={red!0!black},
    urlcolor={blue!0!black}
}

\def\BibTeX{{\rm B\kern-.05em{\sc i\kern-.025em b}\kern-.08em
    T\kern-.1667em\lower.7ex\hbox{E}\kern-.125emX}}
\usepackage[skip=0pt]{caption}
 \newcommand{\Lim}[1]{\raisebox{0.5ex}{\scalebox{0.8}{$\displaystyle \lim_{#1}\;$}}}
\begin{document}
\bstctlcite{IEEEexample:BSTcontrol}
\title{Over-the-Air Federated Learning with Energy Harvesting Devices\\

\thanks{The authors acknowledge support from TUBITAK through CHIST-ERA project SONATA (CHIST-ERA-20-SICT-004, funded by TUBITAK, Turkey Grant 221N366 and EPSRC, UK Grant EP/W035960/1).

Ozan Ayg{\"u}n's research is also supported by Turkcell A.S. within the framework of 5G and Beyond Joint Graduate Support Programme coordinated by Information and Communication Technologies Authority.}
}

\author{\IEEEauthorblockN{Ozan Ayg{\"u}n\textsuperscript{1}, Mohammad Kazemi\textsuperscript{1}, Deniz G{\"u}nd{\"u}z\textsuperscript{2} and Tolga M. Duman\textsuperscript{1}}
\IEEEauthorblockA{\textsuperscript{1}\textit{Dept. of Electrical and Electronics Engineering, Bilkent University}, Ankara, Turkey \\
\textsuperscript{2}\textit{Dept. of Electrical and Electronic Engineering, Imperial College London}, London, UK \\
\{ozan, kazemi, duman\}@ee.bilkent.edu.tr, d.gunduz@imperial.ac.uk}
}
\maketitle
\begin{abstract}
We consider federated edge learning (FEEL) among mobile devices that harvest the required energy from their surroundings, and share their updates with the parameter server (PS) through a shared wireless channel. In particular, we consider energy harvesting FL with over-the-air (OTA) aggregation, where the participating devices perform local computations and wireless transmission only when they have the required energy available, and transmit the local updates simultaneously over the same channel bandwidth. In order to prevent bias among heterogeneous devices, we utilize a weighted averaging with respect to their latest energy arrivals and data cardinalities. We provide a convergence analysis and carry out numerical experiments with different energy arrival profiles, which show that even though the proposed scheme is robust against devices with heterogeneous energy arrivals in error-free scenarios, we observe a 5-10\% performance loss in energy harvesting OTA FL.
\end{abstract}

\begin{IEEEkeywords}
Federated learning, energy harvesting devices, wireless communications, machine learning. 
\end{IEEEkeywords}

\section{Introduction} \label{sec:Introduction}
Developments in Internet-of-things (IoT) devices have helped machine learning (ML) approaches to be used in many domains such as healthcare, automation, and forecasting, thanks to its endless data collection. While mobile devices are at the center of attention for collecting data, traditional ML approaches require the collected data to be assembled in a cloud server for model training. However, this approach may not be feasible due to several reasons. Firstly, the participants are typically reluctant to share their private data; secondly, sending all the data to a server has a high communication cost, particularly in bandwidth and energy-limited scenarios. Finally, latency can be a critical limitation for time-sensitive applications \cite{Lim2020}. \textit{Federated learning} (FL) is a recently emerging framework that aims to mitigate these issues, where the participating devices perform model training with local data and send their parameter updates to the parameter server (PS), which orchestrates the learning process, instead of sharing the local data itself to preserve privacy \cite{Konecny2016, Mcmahan2017}.

In FL, participating devices called mobile devices (MDs) can be selected based on their available energy, computing capability, and channel quality to the PS \cite{Gunduz2020}. Before the local training, the global model is sent by the PS to the MDs. Selected MDs perform stochastic gradient descent (SGD) iterations using their local dataset. After completion, a subset of the MDs shares their model updates with the PS, where the model aggregation is performed to obtain the updated global model. These steps are repeated either for a prescribed number of iterations or until a certain condition is met. Recent studies on FL include investigating the effects of data heterogeneity \cite{Konecny2016, zhang2021client, Sery2021}, design of communication-efficient approaches \cite{Zhu2020, Zhu2021, MohammadiAmiri2020, Chen2021, tegin, tegin2}, and latency and power analysis \cite{Dinh2021, Liu2021}.

Even though FL has many potential benefits in terms of privacy and communication cost, bandwidth limitations and adverse channel conditions in wireless setups may threaten its feasibility in certain practical scenarios. To reduce the required bandwidth in FEEL, over-the-air (OTA) aggregation has become the \textit{de facto} approach where the same bandwidth is shared by all the participating MDs, enabling the aggregation of gradients during the transmission \cite{MohammadiAmiri2020}. The adverse channel effects can be alleviated using multiple receive antennas and combining techniques at the PS \cite{Amiri2021a, Amiri2021b, ren2020scheduling, aygun2022}.

Despite the success of FL in practical scenarios, the energy consumption and carbon footprint of MDs for training and sharing their local models create serious concerns about the sustainability of future smart systems \cite{strubell2019energy}. As a more sustainable approach, energy harvesting devices, which can acquire energy from their surroundings \cite{ulukus2015energy}, have been widely considered for mobile networks. These devices are typically equipped with a rechargeable battery to store the harvested energy, and perform the required computations and communications if they have available energy in their battery. 

Energy harvesting communication devices have been previously studied in detail from different perspectives, including optimal transmission policies \cite{ozel2011transmission, gurakan2013energy, ozel2015fundamental}, and channel capacity computation for unit-sized battery \cite{tutuncuoglu2017binary}. Energy harvesting FL has also been considered where the participating MDs are energy harvesting devices \cite{guler2021energy, hamdi2021federated}. However, no approach considers the wireless channel effects and OTA aggregation in energy harvesting FL setups, which constitutes the basis of this work.

To examine the performance of energy harvesting FL in a practical setting, we introduce energy harvesting OTA FL where the participating MDs perform their local SGD iterations and transmit their gradients using the wireless links whenever enough energy is available. Using OTA aggregation and combining techniques, the PS updates the global model based on the received signal. The updated model is sent back to the devices for the next global iteration. We compare the performance of our setup with the error-free scenarios and conventional FL using different energy arrival profiles. Numerical and experimental results show that our proposed algorithm can perform well under practical channel scenarios with convergence guarantees. 

The rest of the paper is organized as follows. In Section \ref{sec:systemmodel}, we introduce the FL setup as well as the energy harvesting processes at the devices with different energy arrival profiles. In Section \ref{sec:hfl}, we study the OTA communication model of FL with MDs that have intermittent energy arrivals. In Section \ref{sec:convergenceanalysis}, the convergence analysis of energy harvesting FL is presented under certain convexity assumptions on the loss function. We demonstrate the numerical results in Section \ref{sec:simulationresults}, and conclude the paper in Section \ref{sec:conclusions}.
\section{System Model} \label{sec:systemmodel}
\subsection{FL Setup}
The main goal in FL is to minimize the global loss $F(\bm{\theta})$ with respect to the model weights $\bm{\theta} \in \mathbb{R}^{2N}$, where $2N$ is the dimension of the weights in the model. Our system has $M$ single-antenna MDs and a PS equipped with $K$ antennas. Each MD has a dataset $\mathcal{B}_{m}$ with cardinality $| \mathcal{B}_{m}|$, and we define $B \triangleq \sum_{m = 1}^{M} |\mathcal{B}_{m}|$ as the number of total data samples. We define the global loss function as
\begin{equation}
    F(\bm{\theta}) = \sum_{m = 1}^{M} \frac{|\mathcal{B}_{m}|}{B} F_{m}(\bm{\theta}),
    \label{eq:empiricalloss}
\end{equation}
where $F_{m}(\bm{\theta}) \triangleq \frac{1}{|\mathcal{B}_{m}|} \sum_{u \in \mathcal{B}_{m}} f(\bm{\theta},u)$, with $f(\bm{\theta},u)$ corresponding to the loss of the $u$-th data sample.

In every global iteration, the MDs perform $\tau$ local SGD iterations using their local data to obtain model updates that needs to be shared with the PS for the global aggregation. The SGD steps at the $m$-th MD at the $i$-th local and $t$-th global iteration are performed as
\begin{equation} \label{eq:sgd}
    \bm{\theta}_{m}^{i+1}(t) = \bm{\theta}_{m}^{i}(t) - \eta_{m}^{i}(t) \nabla F_{m} (\bm{\theta}_{m}^{i}(t), \bm{\xi}_{m}^{i}(t)),
\end{equation}
where $\eta_{m}^{i}(t)$ is the learning rate, $\nabla F_{m} (\bm{\theta}_{m}^{i}(t), \bm{\xi}_{m}^{i}(t))$ is the unbiased local gradient estimate for the local weights $\bm{\theta}_{m}^{i}(t)$ with the randomly sampled batch $\bm{\xi}_{m}^{i}(t)$ from the dataset $\mathcal{B}_{m}$, i.e., $\mathbb{E}_{\xi} \left[ \nabla F_{m} (\bm{\theta}_{m}(t), \bm{\xi}_{m}(t)) \right] = \nabla F_{m} (\bm{\theta}_{m}(t)),$ where the expectation is over the random batch of data samples.

Having computed the local SGD steps, MDs calculate their model difference to be shared with the PS as
\begin{equation}
    \Delta \bm{\theta}_{m}(t) = \bm{\theta}_{m}^{\tau}(t) - \bm{\theta}_{m}^{1}(t).
\end{equation}
In the case where all the devices are participating in the global aggregation with error-free transmission, the PS performs the global aggregation using
\begin{equation}
    \bm{\theta}_{PS}(t+1) = \bm{\theta}_{PS}(t) + \sum_{m = 1}^{M} p_{m}(t) \Delta \bm{\theta}_{m}(t),
\end{equation}
where $\bm{\theta}_{PS}(t)$ represents the model weight vector at the PS at the $t$-th global iteration and $p_{m}(t) = \frac{|\mathcal{B}_{m}|}{\sum_{m = 1}^{M} |\mathcal{B}_{m}|}$ denotes the ratio of the number of data samples of the $m$-th device to the total number of samples participating in the aggregation. Note that the denominator can change depending on the number of participating devices. The updated global weights at the PS are shared with the MDs for the next global iteration.

Global aggregation can also be performed via OTA aggregation, where the local model updates can be transmitted over a shared wireless medium to the PS, whose output for the $k$-th receive antenna becomes \footnote{A similar setup can also be obtained using orthogonal frequency-division multiplexing (OFDM).}
\begin{equation}
    \bm{y}_{PS,k}(t) = \sum_{m \in \mathcal{S}_{t}} \bm{h}_{m,k}(t) \circ \bm{x}_{m}(t) + \bm{z}_{PS,k}(t),
\end{equation}
where $\bm{x}_{m}(t)$ is the transmitted signal from the $m$-th MD, $\circ$ is the element-wise product, $\bm{z}_{PS,k}(t) \in \mathbb{C}^{N}$ is the circularly symmetric additive white Gaussian noise (AWGN) vector with independent and identically distributed (i.i.d.) entries with zero mean and variance of $\sigma_{z}^{2}$; i.e., $ z_{PS,k}^{n}(t) \sim \mathcal{CN}(0,\sigma_{z}^{2})$. The channel coefficients are given as $\bm{h}_{m,k}(t) = \sqrt{\beta_{m}} ~ \bm{g}_{m,k}(t)$, where $\bm{g}_{m,k}(t) \in \mathbb{C}^{N}$ with each entry $g_{m,k}^{n}(t) \sim \mathcal{CN}(0,\sigma_{h}^{2})$ (i.e., Rayleigh fading), $\beta_{m}$ is the large-scale fading coefficient modeled as $\beta_{m} = \left( d_{m} \right)^{-p}$, where $p$ denotes the path loss exponent, and $d_{m}$ represents the distance between $m$-th MD and the PS. 
%% FL end
\subsection{Energy Harvesting Devices}
We consider energy harvesting MDs, which harvest either unit energy, or no energy at all from various sources such as solar, kinetic, or RF energy in every global iteration. For simplicity, we assume that $\tau$ local SGD steps and the transmission of gradients to the PS cost a unit amount of energy, and each MD has a unit battery.

We denote the binary energy arrival process of the $m$-th MD at the $t$-th global iteration as $E_{m}(t)$. If $E_{m}(t) = 1$, this means that $m$-th MD receives the enough energy to participate in the global iteration at iteration $t$. $E_{m}(t) = 0$, if no energy is harvested. We also define the elapsed time between the current iteration and the previous energy arrival as $\lambda_{m}(t) = \max_{t^{\prime}:t^{\prime}<t, E_{m}(t^{\prime}) = 1} t^{\prime}$. Lastly, for a given $t$, we denote the cooldown multiplier as $c_{m}(t) = t - \lambda_{m}(t),$ which represents for how many iterations the $m$-th MD has not been harvesting energy. 

We investigate MDs with stochastic energy arrival profiles, where the harvested energy has an underlying probability distribution, and the MDs have no prior information about when the next arrival will be. Note that the MDs do not need to know the underlying distribution of the stochastic process. We will be covering the following stochastic energy arrival processes.
\subsubsection{Bernoulli}
At the $t$-th global iteration, the $m$-th MD receives energy with probability $\alpha_{m}(t)$, i.e., 
\begin{equation}
    E_{m}(t) = 
    \begin{cases}
    1 \hspace{0.3cm} \text{with probability} \hspace{0.3cm} \alpha_{m}(t), \\
    0 \hspace{0.3cm} \text{with probability} \hspace{0.3cm} 1-\alpha_{m}(t).
    \end{cases}
\end{equation}
\subsubsection{Uniform}
Global iterations are divided into blocks of $T_{m}$, and the $m$-th MD receives energy once in every $T_{m}$ iterations. This means that with probability 1, an energy arrival is observed within time instances $\big\{ t,\ldots,t + T_{m} - 1 \big\}$. 
% \section{Hierarchical OTA Federated Learning (HOTAFL)} \label{sec:hfl}
\section{OTA FL with Energy Harvesting} \label{sec:hfl}
We now describe the proposed scheme for which FL participants are energy harvesting devices and the gradients are sent through wireless channels using OTA aggregation. 

Devices do not always have sufficient energy to perform the local SGD iterations and gradient transmission, so only the MDs that have harvested enough energy, i.e., $E_{m}(t) \!=\!  1$ can participate in the $t$-th global iteration. We define $\mathcal{S}(t)$ as the set of devices participating in the $t$-th global iteration.

Before each training round, the MDs receive the current global model $\bm{\theta}_{PS}(t)$ from the PS. If an MD is eligible to participate in the $t$-th iteration based on its energy status, the SGD calculations are performed. Then, based on cooldown multiplier of each MD, the weighted model differences are calculated as
\begin{equation}
    \Delta \bm{\theta}_{m}^{s}(t) = C_{m}(t) \Delta \bm{\theta}_{m}(t),
\end{equation}
where $C_{m}(t) = p_{m}(t) c_{m}(t)$, and $\Delta \bm{\theta}_{m}^{s}(t)$ denotes the scaled model differences for the $m$-th MD at the $t$-th global iteration. Considering error-free transmission of the scaled gradients, the PS performs global update for the next iteration as 
\begin{equation}
    \bm{\theta}_{PS}(t+1) = \bm{\theta}_{PS}(t) + \Delta \bm{\theta}_{PS}(t),
\end{equation}
where we define $\Delta \bm{\theta}_{PS}(t)$ as
\begin{equation} \label{eq:delta_theta_ps}
    \Delta \bm{\theta}_{PS}(t) = \frac{1}{C(t)} \sum_{m \in \mathcal{S}_{t}} \Delta \bm{\theta}_{m}^{s}(t)
\end{equation}
with $C(t) = \sum_{m \in \mathcal{S}_{t}} C_{m}(t)$, which is assumed to be known by the PS \cite{guler2021energy}. The readers are referred to \cite{liu2018massive} and the references therein for related algorithms to estimate the number of participating users.
%\subsection{OTA Aggregation}

We now consider the OTA aggregation of the local model differences. The PS receives a noisy target signal due to the wireless channel and the noise. In the proposed scheme, we assume perfect channel state information (CSI) at the receiver side and no CSI at the MDs. 

For a more spectrally efficient approach, the model differences are written in terms of a complex signal $\Delta \bm{\theta}_{m}^{s,cx}(t) \in \mathbb{C}^{N}$ by grouping the symbols into its real and imaginary parts as
\begin{subequations}\label{eq:reco}
    \begin{alignat}{2}
        & \!\!\Delta \bm{\theta}_{m}^{s,re}(t) \triangleq \left[ \Delta \theta_{m}^{s,1}(t), \Delta \theta_{m}^{s,2}(t), \ldots, \Delta \theta_{m}^{s,N}(t) \right]^{T}, \label{sub-eq-1:reco} \\
        & \!\!\Delta \bm{\theta}_{m}^{s,im}(t) \!\triangleq \!\! \left[ \Delta \theta_{m}^{s,N+1}\!(t), \Delta \theta_{m}^{s,N+2}(t), \ldots, \Delta \theta_{m}^{s,2N}\!(t)\! \right]^{\!T}\!\!. \label{sub-eq-2:reco} 
    \end{alignat}
\end{subequations}
For the $k$-th antenna, the PS receives the following signal
\begin{equation}
    \bm{y}_{PS,k}(t) = \sum_{m \in \mathcal{S}_{t}} \bm{h}_{m,k}(t) \circ \Delta \bm{\theta}_{m}^{s,cx}(t) + \bm{z}_{PS,k}(t),
\end{equation}

Since we have the perfect CSI at the receiver side, the combining can be done at the PS as (see \cite{Amiri2021a})
\begin{equation}
    \bm{y}_{PS}(t) \!=\! \frac{1}{K} \!\sum_{k = 1}^{K} \! \Big( \!\sum_{m \in \mathcal{S}_{t}} \bm{h}_{m,k}(t) \Big)^{\!\!\ast} \! \circ \! \bm{y}_{PS,k}(t)    
\end{equation}
For the $n$-th symbol, the combined signal becomes
\begin{alignat}{2}
    y_{PS}^{n}(t) \! & = \! \underbrace{\sum_{m \in \mathcal{S}_{t}} \! \Big( \! \frac{1}{K} \! \sum_{k = 1}^{K} \lvert h_{m,k}^{n}(t) \rvert^{2} \! \Big) \! \Delta \theta_{m,s}^{n,cx}(t)}_{\text{$y_{PS}^{n,sig}(t)$ (signal term)}} \nonumber \\
    & \hspace{0.3cm} +\underbrace{  \frac{1}{K} \! \sum_{m \in \mathcal{S}_{t}} \! \sum_{\substack{m^{\prime} \in \mathcal{S}_{t} \\ m^{\prime} \neq m}} \! \sum_{k = 1}^{K} \! (h_{m,k}^{n}(t))^{\ast} \! h_{m^{\prime},k}^{n}(t)  \Delta \theta_{m^{\prime},s}^{n,cx}(t)}_{\text{$ y_{PS}^{n,int}(t)$ (interference term)}}  \nonumber \\
    & \hspace{0.3cm} +\underbrace{  \frac{1}{K} \! \sum_{m \in \mathcal{S}_{t}} \! \sum_{k = 1}^{K}  (h_{m,k}^{n}(t))^{\ast} z_{PS,k}^{n}(t)}_{\text{$y_{PS}^{n,noise}(t)$ (noise term)}}.
    \label{eq:sin1}
    \end{alignat}
% \end{equation}
We recover the aggregated model differences from the received signal as
\begin{subequations}\label{eq:sin11}
    \begin{alignat}{2}
        \Delta \hat{\theta}_{PS}^{n}(t) &= \frac{1}{C(t)\sigma_{h}^{2} \Bar{\beta}} \operatorname{Re}\{ y_{PS}^{n}(t) \}, \label{sub-eq-1:sin11} \\
        \Delta \hat{\theta}_{PS}^{n+N}(t) &= \frac{1}{ C(t) \sigma_{h}^{2} \Bar{\beta}} \operatorname{Im}\{ y_{PS}^{n}(t) \}. \label{sub-eq-2:sin11}
    \end{alignat}
\end{subequations}
Finally, the global update can be performed as
\begin{equation}
    \bm{\theta}_{PS}(t+1) = \bm{\theta}_{PS}(t) + \Delta \hat{\bm{\theta}}_{PS}(t),
\end{equation}
where $\Delta \bm{\hat{\theta}}_{PS}(t) = \big[ \Delta \hat{\theta}_{PS}^{1}(t) ~ \Delta\hat{\theta}_{PS}^{2}(t) ~ \cdots ~ \Delta\hat{\theta}_{PS}^{2N}(t) \big]^{T}$. 
\section{Convergence Analysis} \label{sec:convergenceanalysis}
We denote the minimum local loss as $F_{m}^{*}$, the optimal weights of the model as $\bm{\theta}^{\ast} \triangleq \argmin_{\bm{\theta}} F(\bm{\theta})$, and the minimum total loss function is given as $F^{*} = F(\bm{\theta}^{*})$. The dataset bias is defined as $\Gamma \triangleq F^{*} - \sum_{m = 1}^{M} p_{m} F_{m}^{*} \geq 0$. Moreover, it is assumed that the learning rate remains unchanged among different MDs, i.e., $\eta_{m}^{i}(t) = \eta(t)$. 
\begin{assumption} \label{assumption2} Squared $l_{2}$ norm of the local stochastic gradients are bounded; i.e.,
\begin{equation}
    \mathbb{E}_{\xi} \Big[  \left\| \nabla F_{m} (\bm{\theta}_{m}(t), \bm{\xi}_{m}(t)) \right\|_{2}^{2} \Big] \leq G^{2},
\end{equation}
which translates to $\forall n \!\in\! [2N]$,
    $\mathbb{E}_{\xi} \!\!\left[  \nabla \!F_{m} (\theta_{m}^{n}, \xi_{m}^{n}(t)) \right] \leq\! G$.
\end{assumption}
\begin{assumption} \label{assumption1}
Local loss functions are assumed to be L-smooth and $\mu$-strongly convex; i.e., $\forall \bm{a},\bm{b} \in \mathbb{R}^{2N}$, $\forall m \in [M]$,
\begin{align}
    \!\!F_{m}(\bm{a}) \!-\! F_{m}(\bm{b}) & \!\leq\! \langle \bm{a} \!-\! \bm{b}, \! \nabla\! F_{m} (\bm{b}) \rangle \!+\! \frac{L}{2} \!\left\| \bm{a} - \bm{b} \right\|_{2}^{2}\!, \\
    \!\!F_{m}(\bm{a}) \!-\! F_{m}(\bm{b}) & \!\geq\! \langle \bm{a} \!-\! \bm{b},  \!\nabla\! F_{m} (\bm{b}) \rangle \!+\! \frac{\mu}{2} \!\left\| \bm{a} - \bm{b} \right\|_{2}^{2}\!.
\end{align}
\end{assumption}
\begin{theorem} \label{theorem1}%%%%%%%%%%%%%%%%%%%%%%% Theorem 1
In energy harvesting OTA FL with Bernoulli energy arrivals $\alpha_{m} = \alpha$ and equal data distribution $p_{m} = p, \forall m \in [M]$, for $0 \leq \eta(t) \leq min\{ 1, \frac{1}{\tau \mu} \}$, we can upper bound the model difference between the global and the optimal weights as
\begin{align}
    & \mathbb{E} \big[ \left\| \bm{\theta}_{PS}(t) - \bm{\theta}^{*} \right\|_{2}^{2} \big] \nonumber \\ 
    & \leq \! \bigg( \prod_{a = 1}^{t-1} \! X(a) \! \bigg) \! \left\| \bm{\theta}_{PS}(0) \! - \! \bm{\theta}^{*} \right\|_{2}^{2} \! + \! \sum_{b = 1}^{t-1} \! Y(b) \! \prod_{a = b+1}^{t-1} \! X(a), \label{ourt1}
\end{align}
where $X(a) = \left( 1 - \mu \eta(a) \left( \tau - \eta(a) (\tau - 1) \right) \right)$ and
\begin{align} 
     Y(a) &\!= \tau^{2} G^{2} \eta^{2}(a) \sum_{m_{1} \in \mathcal{S}_{a}} \sum_{m_{2} \in \mathcal{S}_{a}} \!\!\! A(m_{1},m_{2}) \nonumber \\
     & + \frac{\tau^{2} G^{2} \eta^{2}(a)}{K \Bar{\beta}^{2}} \sum_{m \in \mathcal{S}_{a}} \sum_{\substack{m^{\prime} \in \mathcal{S}_{a} \\ m^{\prime} \neq m}} \!\! \beta_{m} \beta_{m^{\prime}} + \frac{\sigma_{z}^{2} N}{p^{2} K \sigma_{h}^{2}} \sum_{m \in \mathcal{S}_{a}} \frac{\beta_{m}}{\Bar{\beta}^{2}} \nonumber \\
     & + \left( 1 + \mu (1- \eta(t) \right) \eta^{2}(t) G^{2} \frac{\tau (\tau - 1) (2 \tau - 1)}{6} \nonumber \\
     & + \eta^{2}(t) (\tau^{2} + \tau - 1) G^{2} + 2 \eta(t) (\tau - 1) \Gamma. \label{eq:xtyt}
\end{align}
with $A(m_{1},m_{2}) = \Big( 1 \! - \! \frac{\beta_{m_{1}}}{\Bar{\beta}} \! - \! \frac{\beta_{m_{2}}}{\Bar{\beta}} + \frac{ ( M \alpha \! + \! 1 ) (K \! + \!1)\beta_{m_{1}} \beta_{m_{2}}}{M \alpha K \Bar{\beta}^{2}} \Big)$.
\end{theorem}
\begin{IEEEproof} %%%%%%%%%%%%%%%%%%%%%%%% Theorem1 proof
Define an auxiliary variable $\bm{v}(t+1) \! \triangleq \! \bm{\theta}_{PS}(t) \! + \! \Delta \bm{\theta}_{PS}(t)$, where $\Delta \bm{\theta}_{PS}(t)$ is defined in \eqref{eq:delta_theta_ps}. Then, we have
\begin{align}
    &\!\!\left\| \bm{\theta}_{PS}(t\!+\!1) \!-\! \bm{\theta}^{\ast} \right\|_{2}^{2} \nonumber 
    \!=\! \left\| \bm{\theta}_{PS}(t\!+\!1) \!-\! \bm{v}(t\!+\!1) + \bm{v}(t\!+\!1) \!-\! \bm{\theta}^{\ast} \right\|_{2}^{2} \nonumber \\
    &\!\!\quad= \left\| \bm{\theta}_{PS}(t+1) - \bm{v}(t+1) \right\|_{2}^{2} + \left\| \bm{v}(t+1) - \bm{\theta}^{\ast} \right\|_{2}^{2} \nonumber \\
    &\!\!\quad\quad+ 2 \langle \bm{\theta}_{PS}(t+1) - \bm{v}(t+1) , \bm{v}(t+1) - \bm{\theta}^{\ast}  \rangle. \label{eq: theorem1argument}
\end{align}
In the following lemmas, we provide upper bounds for \eqref{eq: theorem1argument}. 
\begin{lemma} \label{convproof1} %%%%%%%%%%%%%%%%%% Lemma1 
$\mathbb{E} \Big[ \big\| \bm{\theta}_{PS}(t+1) - \bm{v}(t+1) \big\|_{2}^{2} \Big]$
\begin{align}
    & \leq \tau^{2} G^{2} \eta^{2}(t) \sum_{m_{1} \in \mathcal{S}_{t}} \sum_{m_{2} \in \mathcal{S}_{t}} \!\!\! A(m_{1},m_{2}) + \frac{\sigma_{z}^{2} N}{p^{2} K \sigma_{h}^{2}} \sum_{m \in \mathcal{S}_{t}} \frac{\beta_{m}}{\Bar{\beta}^{2}} \nonumber \\
    &\quad + \frac{\tau^{2} G^{2} \eta^{2}(t)}{K \Bar{\beta}^{2}} \sum_{m \in \mathcal{S}_{t}} \sum_{\substack{m^{\prime} \in \mathcal{S}_{t} \\ m^{\prime} \neq m}} \beta_{m} \beta_{m^{\prime}}.
\end{align}
\end{lemma}
\begin{IEEEproof} %%%%%%%%%%%%%%%%%% Lemma1 Proof
See Appendix \ref{sec:appendix1}.
\end{IEEEproof}
\begin{lemma} \label{convproof2}
$\!\mathbb{E}\! \Big[ \!\big\| v(t\!\!+\!\!1) \!-\! \bm{\theta}^{*} \!\big\|_{2}^{2} \!\Big]$
\begin{align}
    & \leq\! \!\left( 1 \!-\! \mu \eta(t) \! \left( \tau \!-\! \eta(t) (\!\tau \!\!-\!\! 1) \!\right) \!\right) \!\mathbb{E} \!\Big[ \!\big\| \bm{\theta}_{\!PS}(t) \!-\! \bm{\theta}^{*}  \!\big\|_{2}^{2} \!\Big] \nonumber \\
    & \quad + \left( 1 + \mu (1- \eta(t) \right) \eta^{2}(t) G^{2} \frac{\tau (\tau - 1) (2 \tau - 1)}{6} \nonumber \\
    & \quad + \eta^{2}(t) (\tau^{2} + \tau - 1) G^{2} + 2 \eta(t) (\tau - 1) \Gamma.
\end{align}
\end{lemma}
\begin{IEEEproof}
The proof follows the same line as in Lemma 2 in \cite{Amiri2021a}.
\end{IEEEproof}
\begin{lemma} \label{convproof3}
    $\mathbb{E} \left[ \langle \bm{\theta}_{PS}(t+1) - \bm{v}(t+1) , \bm{v}(t+1) - \bm{\theta}^{\ast}  \rangle \right] = 0$.
    % We have
% \begin{equation}
%     \mathbb{E} \left[ \langle \bm{\theta}_{PS}(t+1) - \bm{v}(t+1) , \bm{v}(t+1) - \bm{\theta}^{\ast}  \rangle \right] = 0.
% \end{equation}
\end{lemma}
\begin{IEEEproof}
The derivation is the same as in Lemma 3 in \cite{aygun2022} by using the independence between local updates and individual channel realizations.
\end{IEEEproof}
The theorem is concluded after applying recursion to the results of Lemmas \ref{convproof1}-\ref{convproof3}. 
\end{IEEEproof}
\begin{corollary}
Using Assumption \ref{assumption1}, the global loss can be upper bounded after $T$ global iterations as
\begin{align} \label{eq:cor1}
    & \!\!\mathbb{E} \left[ F(\bm{\theta}_{PS}(T)) - F^{*} \right] \leq \frac{L}{2} \mathbb{E} \left[ \left\| \bm{\theta}_{PS}(T) - \bm{\theta}^{*} \right\|_{2}^{2} \right] \nonumber \\
    & \!\!\leq \! \frac{L}{2} \! \bigg( \prod_{n = 1}^{T-1} \!\! X(n) \!\! \bigg) \!\! \left\| \bm{\theta}_{PS}(0) \! - \! \bm{\theta}^{*} \right\|_{2}^{2} \! + \! \frac{L}{2} \! \sum_{p = 1}^{T-1} \! Y(p) \!\!\! \!\prod_{n = p+1}^{T-1} \!\!\!\! X(n).
\end{align}
Assuming $\tau = 1, \beta_{m} = 1, \forall m \in [M], \eta(t) = \eta, \forall t$ and knowing that $K \! \gg \! M$, we get
\begin{align}
    & \mathbb{E} \left[ F \big( \bm{\theta}_{PS}(T) \big) \right] - F^{*} \approx \frac{L}{2} \big( 1 - \mu \eta \big)^{T} \! \left\| \bm{\theta}_{PS}(0) \! - \! \bm{\theta}^{*} \right\|_{2}^{2} \nonumber \\
    & \hspace{0.5cm} + \! \frac{L}{2 \mu \eta} \Big( 2 \eta^{2} G^{2} + \frac{\sigma_{z}^{2} N}{p^{2} K \sigma_{h}^{2}} \Big) \Big( 1 - \big( 1 - \mu \eta \big)^{T} \Big). \label{ourc22}
\end{align}
\end{corollary}
\noindent {\textbf{Remark.}} The noise term in $Y(t)$ does not depend on $\eta(t)$, so we have $\Lim{t \to \infty} \mathbb{E} [F(\bm{\theta}_{PS}(t))] - F^{*} \neq 0$ even though $\Lim{t \to \infty} \eta(t) = 0$. As expected, having more receive antennas and more data contribution from devices increases the convergence rate, whereas the model size and the noise variance have negative effects.
\section{Simulation Results} \label{sec:simulationresults}
The experiments are made with an FL environment with $M = 40$ MDs and a PS with $K = 5M$ receive antennas. MDs are spread around the PS randomly such that their distances to the PS is uniformly distributed between 0.5 and 2. 

We use CIFAR-10 \cite{cifar10} dataset with Adam optimizer \cite{adam}, and consider the i.i.d. data distribution where the data samples are randomly and equally distributed among MDs. The same architecture presented in \cite{Amiri2021a} is used with $2N=307498$.

In the simulations, we have considered conventional FL without wireless links, OTA FL where all the MDs have available energy to participate at all iterations, and energy harvesting FL where MDs have intermittent energy arrivals with both error-free and OTA aggregation schemes. To make a comparison with the previous studies, we also consider the setup used in \cite{guler2021energy} with Bernoulli energy arrivals, which corresponds to our error-free energy harvesting FL setup without any normalization at the PS with respect to the cooldown multipliers. Moreover, the MDs are divided into 4 equal-sized groups with different energy profiles. For Bernoulli energy arrivals, we have $\alpha_{m}(t) = \big( 1, 1/5, 1/10, 1/20 \big)$, and for uniform energy arrivals, we have $T_{m} = \big( 1,5,10,20 \big)$ for MDs in 4 groups as in \cite{guler2021energy}. The training is performed for $T = 1000$ global iterations for $\tau = 1$, and $T =400$ for $\tau = 3$ with mini-batch size $|\bm{\xi}_{m,c}^{i}(t)| = 128$, the path loss exponent $p=4$, $\sigma_{h}^{2} = 1$, and $\sigma_{z}^{2} = 1$. 

\begin{figure}[t]
  \centering
  \includegraphics[width=.955\columnwidth]{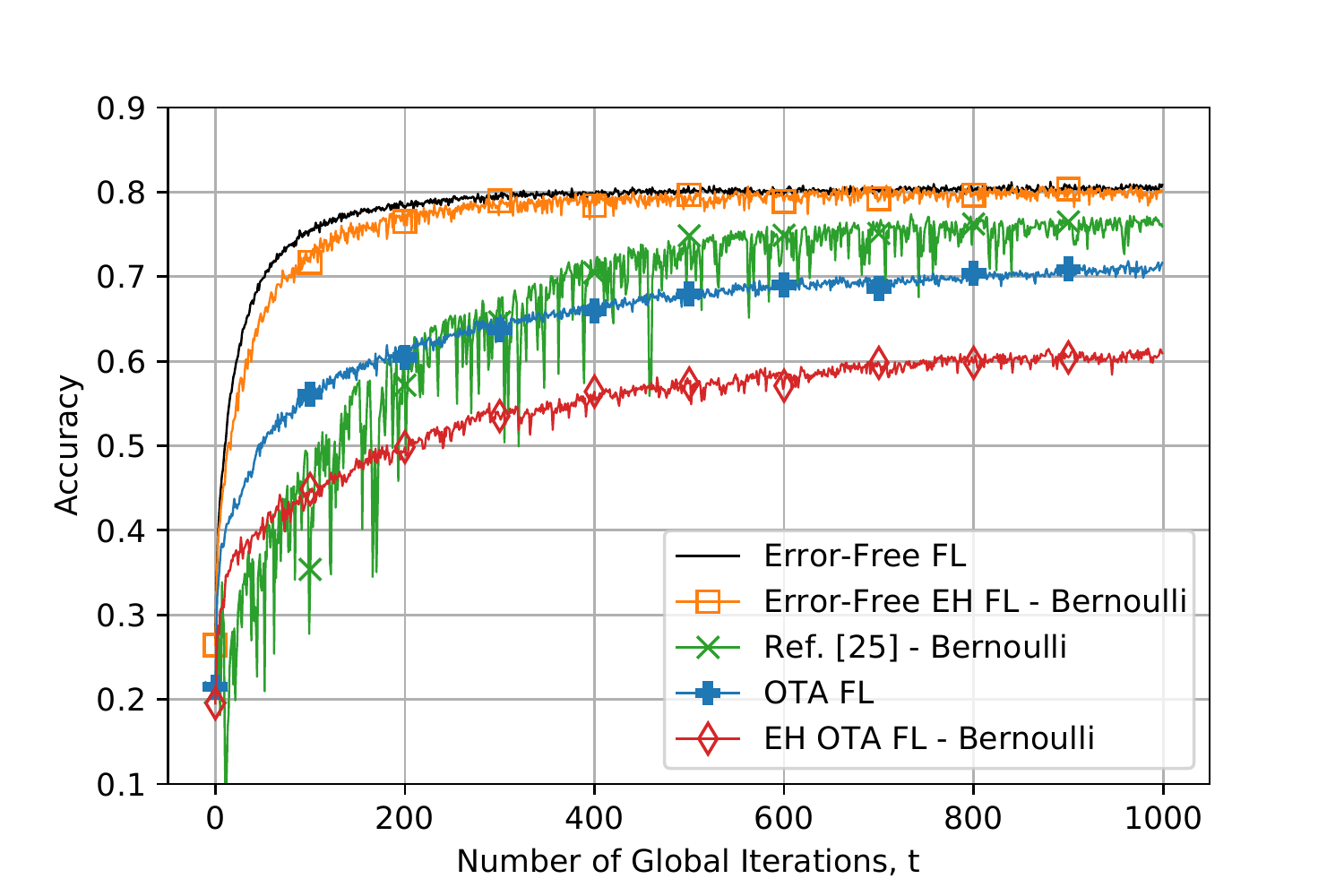} 
\caption{Test accuracy for $\tau = 1$}
\label{fig:cifar10}
\vspace{2mm}
  \centering
  \includegraphics[width=.955\columnwidth]{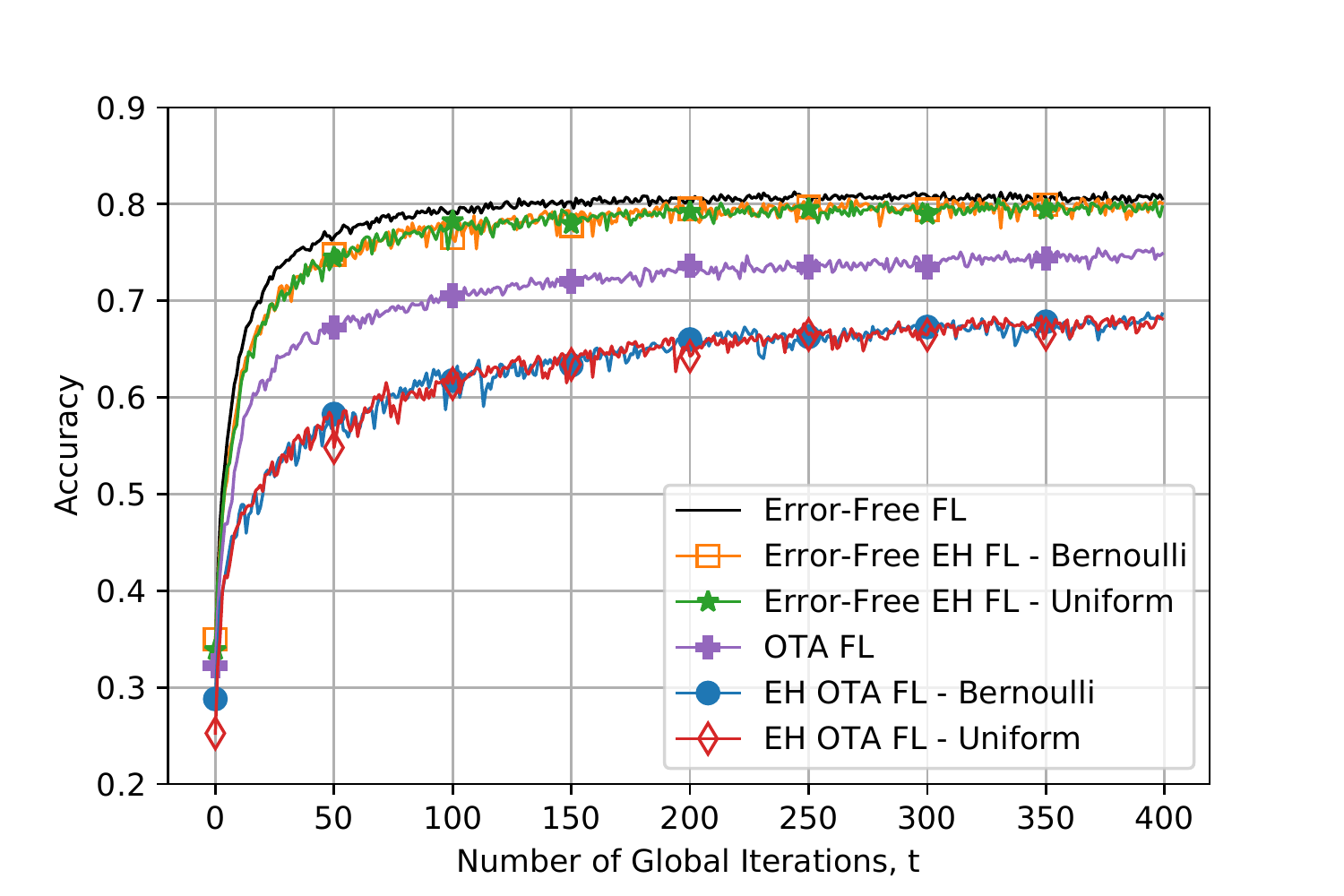} 
\caption{Test accuracy for $\tau = 3$}
\label{fig:cifar10_2}
\vspace{2mm}
  \centering
 \includegraphics[width=.955\columnwidth]{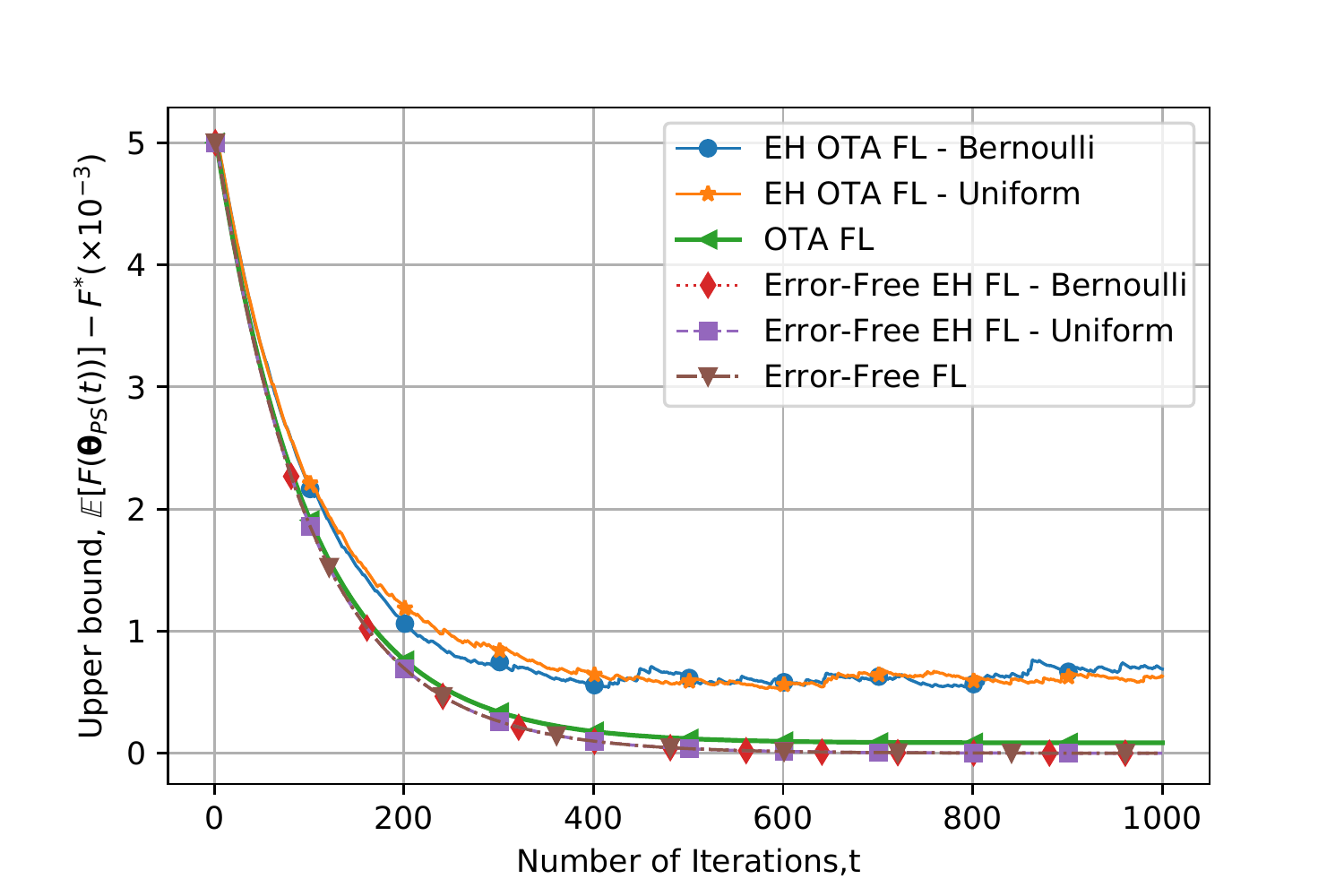} 
\caption{Upper bound on $\mathbb{E} \left[ F(\bm{\theta}_{PS}(t)) - F^{*} \right]$}
\label{fig:convergencerate}
\end{figure}

Accuracy plots for the Bernoulli energy arrival profiles with $\tau \!\! = \!\! 1$ and $\tau \!\! = \!\! 3$ are presented in Figs. \ref{fig:cifar10} and \ref{fig:cifar10_2}, respectively. The results show that the energy harvesting FL with error-free links has a convergence rate close to that of FL with full participation, and that adding a normalization term with respect to the cooldown multipliers leads to a faster convergence and less fluctuations compared to the results in \cite{guler2021energy}. Moreover, OTA FL perform very similar to the scenario used in \cite{guler2021energy} with error-free links. It can be seen that even though the links are wireless, the gap in the performance can be compensated as the number of global iterations increases. One reason is that the increased number of receive antennas at the PS can reduce the adverse affects of the small-scale fading and noise. Increasing $\tau$ achieves a better performance with faster convergence, at the cost of making more computations at the edge. It can also be observed that the performance of Bernoulli arrivals is very close to the that of the uniform arrivals due to the similarities in the energy arrival profiles. 

In Fig. \ref{fig:convergencerate}, we numerically evaluated the convergence rates of the scenarios that we have considered in Fig. \ref{fig:cifar10_2}, using the expression in (\ref{eq:cor1}) with $M \! = \! 40, 2N \!=\! 307498, L \!=\! 10, \mu \!=\! 1, \tau = 1, G^{2} \!=\! 1, \eta(t) \!=\!10^{-2} \! - \!  10^{-6}t, \sigma_{z}^{2} \! = \! 5, \sigma_{h}^{2} = 1, K \! = \! M, \left\| \bm{\theta}_{PS}(0) \!-\! \bm{\theta}^{*} \right\|_{2}^{2} \!=\! 10^{3}$. We observe a close convergence rate between the conventional FL and the error-free energy harvesting FL as expected due to weighted averaging operation with respect to the cooldown multipliers. Energy harvesting FL with OTA aggregation has a slower convergence rate when compared to the others because of the wireless channel as well as the decreased number of participants at each iteration due to energy harvesting devices. We can observe that changing the energy arrival profiles and introducing MDs with less frequent energy arrivals affect $|\mathcal{S}_{t}|$ and $C(t)$, which are key reasons in the shifts and fluctuations in the convergence rates. 
\section{Conclusions} \label{sec:conclusions}
We study OTA FL with energy harvesting devices with intermittent and heterogeneous energy arrivals. Our framework consists of local SGD computations at the MDs that have available energy, and OTA aggregation of the gradients over a shared wireless medium. The convergence rate of FL with energy harvesting devices are examined and its performance is analyzed through numerical experiments. The results with different energy profiles demonstrate that performing a weighted averaging using the latest energy arrival and dataset cardinality in energy harvesting FL can give a similar performance to the full-participation scheme in both error-free and OTA cases. As a future direction, one can investigate different battery capacities, and try to optimize how much power should be allocated for computation and transmission.
\appendices
\section{} 
\label{sec:appendix1}
We can write $\Delta \hat{\theta}_{PS}^{n}(t) = \sum_{p = 1}^{3} \Delta \hat{\theta}_{PS,p}^{n}(t)$, for the $n$-th symbol using \eqref{eq:sin1}, because of the i.i.d. of channel realizations, we obtain
\begin{align}
    &\!\mathbb{E} \big[ || \bm{\theta}_{PS}(t\!+\!1) \!-\! \bm{v}(t\!+\!1) ||_{2}^{2} \big] 
      = \mathbb{E} \big[ \big\| \Delta \hat{\bm{\theta}}_{PS}(t)  \!-\! \Delta \bm{\theta}_{PS}(t) \big\|_{2}^{2} \big] \nonumber\\
    % &  \!\!= \sum_{n = 1}^{2N} \mathbb{E} \big[ ( \Delta \hat{\theta}_{PS}^{n}(t)  - \Delta \theta_{PS}^{n}(t) )^{2} \big]  \nonumber\\
    & \!\!= \!\!\sum_{n = 1}^{2N} (\mathbb{E} \big[ \!\big(\! \Delta \hat{\theta}_{PS,1}^{n}(t)  \!-\! \Delta \theta_{PS}^{n}(t) \big)^{\!2} \big] \!\!+\!\! \sum_{p = 2}^{3} \mathbb{E} \big[ \!\big( \!\Delta \hat{\theta}_{PS,l}^{n}(t) \big)^{\!2} \big]\!.
\end{align}
%%%%%%%%%%%%%%% Term1
\begin{lemma} \label{convproof11} $\mathlarger{\sum_{n = 1}^{2N}} \mathbb{E} \big[ \big( \Delta \hat{\theta}_{PS,1}^{n}(t) \!\!  -  \!\! \Delta \theta_{PS}^{n}(t) \big)^{2} \big]$
\begin{align}
    \hspace{0.1cm} \leq \! \sum_{n = 1}^{2N} \sum_{m_{1} \in \mathcal{S}_{t}} \sum_{m_{2} \in \mathcal{S}_{t}} \!\!\! A(m_{1},m_{2}) \mathbb{E} \big[ \Delta \theta_{m_{1}}^{n}(t) \Delta \theta_{m_{2}}^{n}(t) \big].
    \label{lemma4}
\end{align}
where 
   $A(m_{1},m_{2}) = \Big( 1 \! - \! \frac{\beta_{m_{1}}}{\Bar{\beta}} \! - \! \frac{\beta_{m_{2}}}{\Bar{\beta}} + \frac{ 2 + ( M \alpha \! - \! 1 ) (K \! - \!1)\beta_{m_{1}} \beta_{m_{2}}}{M \alpha K \Bar{\beta}^{2} } \Big)$.
\end{lemma}
\begin{IEEEproof} %%%%%%%%%%%%%% Term1 Proof
For a single symbol, we can write
\begin{align}
    &\mathbb{E} \big[ \big( \Delta \hat{\theta}_{PS,1}^{n}(t)  - \Delta \theta_{PS}^{n}(t) \big)^{2} \big] \nonumber \\
    & = \mathbb{E} \Big[ \frac{1}{C(t)^2} \sum_{m_{1} \in \mathcal{S}_{t}} \sum_{m_{2} \in \mathcal{S}_{t}} C_{m_{1}}(t) C_{m_{2}}(t) \Delta \theta_{m_{1}}^{n}(t)  \Delta \theta_{m_{2}}^{n}(t)  \Big. \nonumber \\ 
    & \quad \Big. \times  \Big( 1 \! - \! \frac{1}{K \sigma_{h}^{2} \Bar{\beta}} \sum_{k_{1} = 1}^{K} |h_{m_{1},k_{1}}^{n}(t)|^{2} \! - \! \frac{1}{K \sigma_{h}^{2} \Bar{\beta}} \sum_{k_{2} = 1}^{K} |h_{m_{2},k_{2}}^{n}(t)|^{2} \Big. \Big. \nonumber \\
    & \hspace{0.8cm} \Big. \Big. + \! \frac{1}{K^{2} \sigma_{h}^{4} \Bar{\beta}^{2}} \!\! \sum_{k_{1} = 1}^{K} \! \sum_{k_{2} = 1}^{K} \!\! |h_{m_{1},k_{1}}^{n}\!(t)|^{2}  |h_{m_{2},k_{2}}^{n}\!(t)|^{2} \! \Big) \! \Big]\!. \label{eq:lemma4expanded}
\end{align}
Using $C_{m}(t) \leq p$ and $C^{2}(t) \leq p^{2}$ and utilizing the i.i.d. channel realizations result in \eqref{lemma4}.
\end{IEEEproof}
\begin{lemma} \label{convproof12} %%%%%%%%%%%%% Term2 Proof
$\mathlarger{\sum_{n = 1}^{2N}} \! \mathbb{E} \big[ \! \big( \! \Delta \hat{\theta}_{PS,2}^{n}(t) \!  \big)^{2} \! \big] \!\!\! \leq \!\!\! \mathlarger{\sum_{m \in \mathcal{S}_{t}} \! \sum_{\substack{m^{\prime} \in \mathcal{S}_{t} \\ m^{\prime} \neq m}}} \!\!\!\! \frac{\beta_{m} \beta_{m^{\prime}}}{K \Bar{\beta}^{2}} \! \mathbb{E} \! \big[ \!\! \left\| \! \Delta \bm{\theta}_{m^{\prime}} \! \big(  t \big) \! \right\|_{2}^{2} \!\! \big]. $
\end{lemma}
\begin{IEEEproof}
For the real part, using the independence of channels for different $m$'s and $k$'s, we obtain
\begin{align}
    &\mathbb{E} \big[ \big( \Delta \hat{\theta}_{PS,2}^{n}(t) \big)^{2} \big] =  \mathbb{E} \Big[ \Big( \sum_{m \in \mathcal{S}_{t}} \sum_{\substack{m^{\prime} \in \mathcal{S}_{t} \\ m^{\prime} \neq m}} \frac{1}{C(t) \sigma_{h}^{2} \Bar{\beta}} \Big. \Big. \nonumber \\
    & \hspace{0.5cm} \Big. \Big. \times \sum_{k = 1}^{K} \operatorname{Re} \big\{ \big(  h_{m,k}^{n}(t) \big)^{*} h_{m^{\prime},k}^{n}(t) C_{m^{\prime}}(t) \Delta \theta_{m^{\prime},c}^{n}(t) \big\} \Big)^{2} \Big] \nonumber\\
    & \leq \mathbb{E} \Big[ \sum_{m \in \mathcal{S}_{t}} \sum_{\substack{m^{\prime} \in \mathcal{S}_{t} \\ m^{\prime} \neq m}} \!\! \frac{\beta_{m} \beta_{m^{\prime}}}{2 K \Bar{\beta}^{2}} \big( \! \big( \Delta \theta_{m^{\prime},c}^{n}(t) \big)^{2} \!\! + \! \big( \Delta \theta_{m^{\prime}}^{n+N}(t) \big)^{2} \nonumber \\
    & \hspace{1cm}+ \Delta \theta_{m}^{n}(t) \Delta \theta_{m^{\prime}}^{n}(t) - \Delta \theta_{m}^{n+N}(t) \Delta \theta_{m^{\prime}}^{n+N}(t) \big) \Big]
\end{align}
% .  
We obtain a similar expression for $N+1 \leq n \leq 2N$, and summing the two parts concludes the lemma.
\end{IEEEproof}
\begin{lemma} \label{convproof13} %%%%%%%%%%%% Term3 Proof
$\mathlarger{\sum_{n = 1}^{2N} \! \mathbb{E}  \big[  \big(  \Delta \hat{\theta}_{PS,3}^{n}(t)  \big)^{2}  \big] \! \leq \! \frac{\sigma_{z}^{2} N}{p^{2} K \sigma_{h}^{2}} \sum_{m \in \mathcal{S}_{t}} \frac{\beta_{m}}{\Bar{\beta}^{2}}}.$
\end{lemma}
\begin{IEEEproof} The first half of the signal yields to
\begin{align} 
    & \mathbb{E} \big[ \big( \Delta \hat{\theta}_{PS,3}^{n}(t) \big)^{2} \big] \nonumber \\
    & = \mathbb{E} \Big[ \Big( \sum_{m \in \mathcal{S}_{t}} \sum_{k = 1}^{K} \frac{1}{C(t) K \sigma_{h}^{2} \Bar{\beta}} \operatorname{Re} \big\{ \big( h_{m,k}^{n}(t) \big)^{*} z_{PS,k}^{n}(t) \big\} \Big)^{2} \Big] \nonumber \\
    & \leq \frac{1}{ p^{2} K^2 \sigma_{h}^{4} \Bar{\beta}^{2}} \mathbb{E} \Big[ \sum_{m \in \mathcal{S}_{t}} \sum_{k = 1}^{K} \big( \operatorname{Re} \big\{ \big( h_{m,k}^{n}(t) \big)^{*} z_{PS,k}^{i,n}(t) \big\} \big)^{2} \Big] \nonumber \\
    & \overset{(a)}{=} \frac{\sigma_{z}^{2}}{2 p^{2} K \sigma_{h}^{2}} \sum_{m \in \mathcal{S}_{t}} \frac{\beta_{m}}{\Bar{\beta}^{2}}.
\end{align}
where (a) is obtained using the independence between the channel realizations and the noise. The result also holds for $N+1 \leq n \leq 2N$. Summing with respect to all symbols completes the proof.
\end{IEEEproof}
The proof is completed using Assumption \ref{assumption2} and \eqref{eq:sgd}, and summing the results in Lemmas \ref{convproof11}-\ref{convproof13}.

\bibliographystyle{IEEEtran}  
\bibliography{main}  

\end{document}